\documentclass[aps,pre,twocolumn,a4paper]{revtex4}
\usepackage{amsmath}
\usepackage{amsfonts}
\usepackage{graphicx}

\DeclareMathOperator{\Var}{Var}

\DeclareMathOperator{\Kurt}{Kurt}

\newcommand{\rv}[1]{\hat{#1}}
\newcommand{\floor}[1]{\left\lfloor#1\right\rfloor}
\newcommand{\ceil}[1]{\left\lceil#1\right\rceil}
\newcommand{\ev}{\mathbb{E}}

\begin{document}

\title{Exact and approximate methods of calculating the sum of states for noninteracting classical and quantum particles occupying a finite number of modes}
\author{Agnieszka Werpachowska}
\email{a.werpachowska@ucl.ac.uk}
\affiliation{Department of Physics and Astronomy, University College London, Gower Street, London WC1E 6BT, United Kingdom}

\begin{abstract}
We present exact expressions for the sum of states of noninteracting classical and quantum particles occupying a finite number of modes with arbitrary spacings. Exploiting a probabilistic analogy, we derive an analytic fourth-order approximation to the density of states, which captures its variance and kurtosis, and is superior to the previous, commonly used methods for all three particle statistics. Our approach employs a simple exact method of calculating the moments of the microcanonical density of states for quantum particles, which requires less computational effort than the commonly used saddle-point approximation. We test our methods numerically and discuss their applicability to various physical systems.
\end{abstract}

\maketitle

\section{Introduction}

Erwin Schr\"{o}dinger wrote ``There is, essentially, only one problem in statistical thermodynamics: the distribution of a given amount of energy $E$ over $N$ identical systems''~\cite{schroedinger}. As the number of particles in a system grows to infinity, almost all its copies (or, expressing the same notion differently, almost all its configurations) have the same internal energy. In this limit, the canonical ensemble (where we control the temperature) is equivalent to the microcanonical ensemble (where we control the energy). The opportunity to pass between these different descriptions gives the possibility to choose the smarter way to measure a given quantity. Very often the calculation of thermodynamical quantities in the microcanonical ensemble is an impractical task and thus one is forced to resort to the canonical ensemble. However, their equivalence breaks down for small or perfectly isolated systems. One then has to work with the microcanonical ensemble, whose workhorse is the sum of states function, which counts the number of microstates realizing a particular value of the control parameter. This is the case, for example, when investigating atoms in microcavities~\cite{philippe:2001} or elongated magneto-optical traps~\cite{bec1d}, heavy nuclei~\cite{Bethe:1936,agrawal:1997}, coupled spins on a lattice~\cite{rilwen} or calculating the entropy of a black hole~\cite{blackholes,Bhaduri:2004}.

In all physical problems listed above the system is composed of $N$ classical (distinguishable), bosonic or fermionic particles which can occupy $S+1$ modes, numbered from 0 to $S$. Each mode can be $g_s$-degenerate and has the excitation $E_s > 0$, with $E_s < E_{s+1}$ and $E_0 = 0$. The last assumption does not lead to the loss of generality, because we can always shift the excitations up or down to ensure that the lowest mode has zero excitation. In many applications, we have $E_s = s$ (e.g.~lattice spins or harmonic traps) or $E_s = s^2$ (square potential wells). 

The control parameter for which we calculate the sum of states is defined as
\begin{equation}
\label{eq:M-sum-ns}
M = \sum_{s=0}^S E_s \sum_{t=0}^{g_s-1} n_{st} \,,
\end{equation}
where $n_{st} = 0, 1, \dotsc$ is the number of particles occupying mode $s$ and (in the case of degenerate modes) its submode $t$, subject to the constraint
\[
\sum_{s = 0}^S \sum_{t=0}^{g_s-1} n_{st} = N \,.
\]
The number of physical states of $N$ particles realizing a given $M$ is called the sum of states $\Omega(N,M,S)$.
When particles are distinguishable, $M$ can be the total magnetization of a system of lattice spins~\cite{rilwen} or the total area of a black hole~\cite{blackholes}. For bosons, $M$ is often the total energy of particles in a harmonic oscillator potential~\cite{bec1d,grossmann} or atoms in a microwave cavity~\cite{philippe:2001}. For fermions, it can be the $z$ component of the total angular momentum of a heavy nucleus~\cite{Bethe:1936,agrawal:1997}. For the sake of generality, we call $M$ a total excitation. 

The paper concerns with the calculation of $\Omega(N,M,S)$ for noninteracting classical, bosonic or fermionic particles occupying a finite number of modes with arbitrary spacings. First, we investigate exact (the popular
recursive and proposed iterative) expressions for the marginal or cumulative sum of states, which are useful in practical calculations. Exploiting a probabilistic analogy, we then obtain an analytic fourth-order approximation to the normalized sum of states (density of states) $\omega$, which captures its mean, variance and kurtosis. The approximation is extended to an analytic-numerical scheme which can also match the skewness of $\omega$ and is better suited to handle excitation spectra with strongly non-homogeneous densities. We demonstrate numerically that our analytic approximation is superior to the previous ones (the Gaussian one and its polynomial corrections fitted also to the kurtosis---e.g.~\cite{Bethe:1936,gram-charlier,agrawal:1997,Blinnikov:1998}) and matches closely the exact sum of states for all three particle statistics. Additionally, our approach employs a simple \textit{exact} method of calculating the moments of the microcanonical density of states for quantum particles, which requires less computational effort than the commonly used saddle-point approximation (see e.g.~\cite{lax:1955,agrawal:1997}). We discuss the applications of our methods to different physical systems pointing out their advantages over commonly employed approaches.

\section{Exact expressions for $\Omega$}

We begin by reviewing the known recursive exact expressions for the sum of states of classical and quantum noninteracting particles~\cite{khinchin:1998}, and proceed later to derive their iterative counterparts.

\subsection{Recursive expressions}

\subsubsection{Sum of states}

The total excitation of $N$ particles distributed among $S+1$ modes is defined by~\eqref{eq:M-sum-ns}. If the mode $S$ (degenerate or not) is occupied by $n$ particles, than the remaining excitation $M - n E_S$ can be redistributed among the $N-n$ particles in $w(N,n) \Omega(N-n,M-nE_S,S-1)$ ways, where the factor 
\[
w(N,n) = \begin{cases}
\binom{N}{n} = \frac{N!}{(N-n)!n!} & \text{(classical)} \\
1 & \text{(quantum)}
\end{cases}
\]
accounts for the particle statistics. On the other hand, if the mode $S$ is $g_S$-degenerate, the $n$ particles occupying it can be reshuffled in
\[
v(g_S,n) = \begin{cases}
(g_S)^n & \text{(classical)} \\
\binom{n + g_S - 1}{g_S - 1} & \text{(bosonic)} \\
\binom{g_S}{n} 1_{n\le g_S} & \text{(fermionic)}
\end{cases}
\]
ways~\cite{Huang:1987_B}. Hence, we have the recursive expression
\begin{equation}
\label{eq:OmgRec1}
\begin{split}
&\Omega(N,M,S) =\\
&\ \sum_{n=0}^{\eta(g_S,N)} w(N,n)v(g_S,n) \Omega(N-n,M-nE_S,S-1) 
\end{split}
\end{equation}
where
\[
\eta(g_S,N) = \begin{cases}
N & \text{(classical/bosonic)} \\
\min(g_S,N) & \text{(fermionic)}
\end{cases} \,.
\]
The boundary condition for $\Omega$ is 
\begin{equation}
\label{eq:OmgBndCond}
\Omega(N,M,0) = \delta_{M,0} v(g_0,N) \,.
\end{equation}

For many values of $n$ the factor $\Omega(N-n,M-nE_S,S-1)$ will be zero, because $n$ will be either too high or too low. Thus, it makes sense to derive tighter bounds for the range of summation over $n$. For its given value $M$, the number $n$ of particles occupying mode $s$ cannot be higher than
\[
Q_S = \min(\floor{M/E_S}, \eta(g_S,N)) \,
\]
(where $\floor{x}$ is the highest integer number less than or equal to $x$), otherwise the total excitation would exceed $M$. 

On the other hand, for classical particles or bosons the maximum $M$ that $N - n$ particles occupying modes below $S$ can realize is $(N-n)E_{S-1}$, which leads to the inequality
\begin{equation}
\label{eq:MineqBCl}
M \le n E_S + (N-n)E_{S-1} = N E_{S-1} + n (E_S - E_{S-1}) \,.
\end{equation}
Since $E_S > E_{S-1}$, $n$ is greater or equal to $P_S$ defined as
\begin{equation}
\label{eq:PSBCl}
P_S = \max \left( \ceil{ \frac{M - N E_{S-1}}{E_S - E_{S-1}} }, 0 \right) \ \text{(class./bosonic)}
\end{equation}
where $\ceil{x}$ is the lowest integer number greater than or equal to $x$. Equation~\eqref{eq:OmgRec1} is modified to
\begin{equation}
\label{eq:OmgRec2}
\begin{split}
&\Omega(N,M,S) =\\
&\ \sum_{n=P_S}^{Q_S} w(N,n)v(g_S,n) \Omega(N-n,M-nE_S,S-1) \,.
\end{split}
\end{equation}

The remaining question is the definition of $P_S$ for fermions, for which it should be higher due to the Pauli exclusion principle. In principle, we can derive an inequality similar to~\eqref{eq:MineqBCl}, but it would be tedious to use, as it would depend on the excitation values of multiple modes $s<S$. However, we can safely use the $P_S$ as defined for bosons and classical particles~\eqref{eq:PSBCl} when dealing with fermions, at the cost of summing over a larger number of zero values.

For classical particles, we can also use recursion in the number of particles, not in the number of modes, writing
\begin{equation}
\label{eq:OmgRec3}
\Omega(N,M,S) = \sum_{s=0}^S g_s \Omega(N-1,M-E_s, S) \,,
\end{equation}
with the same boundary condition as before. This formula holds an advantage over~\eqref{eq:OmgRec2}, because it does not require the evaluation of factorials.

\subsubsection{Cumulative sum of states}

We define the cumulative sum of states as the number of states with total excitation less than or equal to $M$, $\Sigma(N,M,S)$. To distinguish between the two quantities, the previously defined sum of states $\Omega(N,M,S)$ can also be called the \emph{marginal} sum of states. $\Sigma(N,M,S)$ satisfies a recursive equation similar to~\eqref{eq:OmgRec2}, but without the lower limit $P_S$ (which ensured that the total excitation constraint~\eqref{eq:M-sum-ns} was satisfied),
\[
\begin{split}
&\Sigma(N,M,S) =\\
&\ \sum_{n=0}^{Q_S} w(N,n)v(g_S,n) \Sigma(N-n,M-nE_S,S-1) \,,
\end{split}
\]
with a boundary condition also somewhat different from~\eqref{eq:OmgBndCond},
\begin{equation}
\label{eq:SigmaBCond}
\Sigma(N,M,0) = v(g_0,N) \,.
\end{equation}

Analogously to~\eqref{eq:OmgRec3}, we obtain for classical particles only
\[
\Sigma(N,M,S) = \sum_{s=0}^S g_s \Sigma(N-1,M-E_s, S) \,,
\]
with the same boundary condition~\eqref{eq:SigmaBCond}.

In numerical computations, the recursive formulas~\eqref{eq:OmgRec2} and~\eqref{eq:OmgRec3} require the storage of the intermediate values of $\Omega$ or $\Sigma$. This means that the excitation values $E_s$ and $M$ must be discretized, which for irregular excitation values leads to large memory requirements. On the other hand, they do not iterate over every configuration of the system, avoiding the problem of the exponential growth of their total number.

\subsection{Iterative expressions}

\subsubsection{Sum of states}

In certain applications, it is necessary to have access to each of the system configurations counted by the sum of states. For this purpose, the recursive formulas can be converted to an explicit iterative expression for the sum of states:
\begin{equation}
\label{eq:OmgExpl}
\Omega(N,M,S) = \sum_{n_{S} = P_{S}}^{Q_{S}} \dotsm \sum_{n_1 =
P_1}^{Q_1} u(\{ n_s \}) \,,
\end{equation}
where $n_s = \sum_{s'=0}^{g_s-1} n_{ss'}$,
\begin{align*}
P_s &= \max\left(
\ceil{ \frac{ M -E_{s-1}N - \sum_{t=s+1}^{S} (E_t - E_{s-1}) n_t }{E_s - E_{s-1}} } , 0 \right) ,\\
Q_s &= \min\left( \floor{ \frac{M - \sum_{t = s+1}^S E_t n_t}{E_s} }, \eta(g_s, N - \sum_{t=s+1}^S n_t) \right) ,
\end{align*}
and $n_0 = N - \sum_{t = 1}^S n_t$.

The factor 
\[
u(\{ n_s \}) = \prod_{s=0}^S w\left(N - \sum_{t=s+1}^S n_s, n_s\right) v(g_s,n_s)
\]
is the number of particle states realizing a given combination $\{ n_s \}$. Equation~\eqref{eq:OmgExpl} provides a simple way to sum over all combinations of $\{n_{s}\}$, subject to the constraints $\sum_{s=0}^{S} n_{s} = N$ and $\sum_{s=0}^{S} n_{s} E_s = M$, and takes into account the mode degeneracies. Again, the difference between the formulas for bosons and fermions amounts to excluding, in the latter case, all combinations where at least one $n_s > g_s$.

\subsubsection{Cumulative sum of states}

The difference between $\Omega(N,M,S)$ and $\Sigma(N,M,S)$ is the relaxation of constraint~\eqref{eq:M-sum-ns} when calculating the latter. Hence, we can easily write an iterative exact formula for it, similar to~\eqref{eq:OmgExpl},
\[
\Sigma(N,M,S) = \sum_{n_{S} = 0}^{Q_{S}} \dotsm \sum_{n_1 =
0}^{Q_1} u(\{ n_s \}) \,.
\]
The other conditions and definitions remain unchanged.

\section{Analytic approximations for $\Omega$}

Formulas for the sum of states~\eqref{eq:OmgRec2},~\eqref{eq:OmgRec3} and~\eqref{eq:OmgExpl}, though accurate, are not very convenient for analytic calculations. We will now derive two analytic approximations for $\Omega(N,M,S)$ (more precisely, for its version $\omega(N,M,S)$ normalized to unity), by exploiting an analogy between a system of classical/quantum independent particles and a set of independent/correlated random variables.

\subsection{Classical particles}
\label{sec:class-part}

When dealing with classical particles, one can associate the excitation state of each particle with an independent, uniformly distributed random variable $\rv{M}_j$ taking real values $E_0,\dotsc,E_S$, each having a probability $1/\sum_{s=0}^S g_s$, so that its expected value $\ev[\rv{M}_j] = \sum_{s=0}^S g_s E_s / \sum_{s=0}^S g_s$ and variance
\[
\Var[\rv{M}_j] = \ev[(\rv{M}_j - \ev[\rv{M}_j])^2] = \frac{\sum_{s=0}^{S} g_s (E_s - \ev[\rv{M}_j])^2}{\sum_{s=0}^S g_s}  \,.
\]
The total excitation $M$ corresponds to the sum of these variables, $\rv{M} = \sum_{j=1}^N \rv{M}_j$ and the probability that $\rv{M} = M$ is given by $P(\rv{M}=M) = \omega(N,M,S)$. The variance of $\rv M$ is thus
\begin{equation}
\label{eq:varM}
\Var[\rv{M}] = \sum_{j = 1}^N \Var[\rv M_j]
\end{equation}
and the mean
\begin{equation}
\label{eq:expM}
\ev[\rv M] = \sum_{j=1}^N \ev[\rv{M}_j] \,.
\end{equation}
In the case of equal mode spacings, $E_s = s$, and $g_s \equiv 1$ we have
\[
\ev[\rv M] = \frac{NS}{2} \quad \text{and} \quad \Var[\rv{M}] = \frac{NS(S+1)}{12} \,.
\]

The numerical examples presented below correspond to this case, but the presented method will be applicable to any reasonably homogeneous excitation spectrum (i.e.~in which the mode spacings do not systematically increase or decrease over the range of modes considered, leading to the skewness of the sum of states function). (We will discuss the non-homogeneous spectra in Sec.~\ref{sec:non-homogeneous}.)

From the Central Limit Theorem it follows that the probability distribution of $\rv M$ is well approximated for large $N$ by a Gaussian distribution $\Phi_{\mu,\sigma^2}$ with mean $\mu$ given by~\eqref{eq:expM} and variance $\sigma^2$ by~\eqref{eq:varM}. Hence,
\[
\frac{\Omega(N,M,S)}{(S+1)^N} \approx \frac{1}{\sqrt{2\pi \Var[\rv{M}]}} \exp \left( - \frac{(M-\ev[\rv M])^2}{2\Var[\rv{M}]}
\right)
\]
which can be also expressed as
\begin{equation}
\label{eq:OmgApprox}
\Omega(N,M,S) \approx \Omega(N,\ev[\rv M],S) \exp \left( - \frac{(M-\ev[\rv M])^2}{2\Var[\rv{M}]}
\right) .
\end{equation}

\begin{figure}[htbp]
\centering\includegraphics[width=\linewidth]{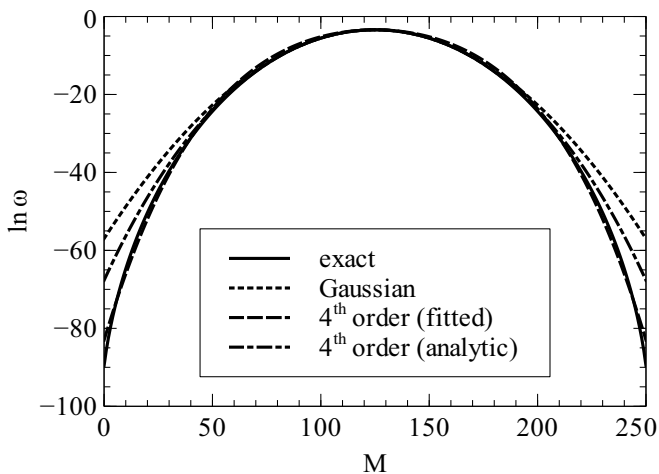}
\caption{Logarithms of exact and approximate values of $\omega(N,M,S)$ for $N = 50$ and $S=5$ (classical particles, equal mode spacings).}
\label{fig:omg}
\end{figure}

In Fig.~\ref{fig:omg} we compare the exact, obtained using the methods from the previous section, and approximate results for $\Omega(N,M,S)$, plotting the logarithm of $\omega(N,M,S)$ to highlight better the discrepancies for extreme values of $M$.
There is an excellent agreement between the Gaussian approximation and exact results around $M=NS/2$. However, further away from the mean their logarithms differ, suggesting that the higher moments of the exact density of states are different from the Gaussian one. Indeed, the excess kurtosis of a single particle variable $\rv{M}_j$ is equal (for $E_s=s$ and $g_s \equiv 1$) to
\[
\Kurt[\rv{M}_j] = \frac{\ev[(\rv{M}_j-\ev[\rv{M}_j])^4]}{\Var[S_j]^2} - 3 = - \frac{12 + 6S(S+2)}{5S(S+2)} \,,
\]
hence
\begin{equation}
\label{eq:kurtM}
\Kurt[\rv M] = \frac{1}{N} \Kurt[\rv{M}_j] = - \frac{12 + 6S(S+2)}{5NS(S+2)} \,.
\end{equation}
The exact probability distribution of $M$ as a random variable is platykurtic (has negative excess kurtosis, i.e.~thin tails), as opposed to the Gaussian distribution, which has zero kurtosis. For large $N$, $\Kurt[\rv M] \to 0$, which is consistent with its distribution approaching Gaussian. For smaller $N$, such as 50, the negative excess kurtosis is still non-negligible and needs to be accounted for. A much better fit to the exact sum of states is obtained by adding a $(M-\ev[\rv M])^4$ term to the exponent of~\eqref{eq:OmgApprox}, yielding $\Omega(N,M,S) \sim \exp( - a (M-\ev[\rv M])^2 -b (M-\ev[\rv M])^4 )$. A numerical least-squares fit of this form to exact data is displayed in Fig.~\ref{fig:omg} as ``4th order (fitted)'' curve.

To derive the analytic approximation including fourth-order terms, we study the properties of the following probability density
\begin{equation}
\label{eq:ford}
\varphi(z) = \frac{2\sqrt{2a}}{e^{1/32a} \sigma K_{1/4}(1/32a)} \exp \left( - \frac{z^2}{2\sigma^2} - \frac{a x^4}{\sigma^4} \right) \,,
\end{equation}
where $K_n(x)$ is the modified Bessel function of the second kind. Our model for $\omega(N,M,S)$ is therefore $\omega(N,M,S) = \varphi(M-\ev[\rv M])$. The case $a = 0$ corresponds to the Gaussian density. For $N = 50$ and $S = 5$, the best-fit value of $a$ (displayed in Fig.~\ref{fig:omg}) is 0.004933, thus we will concentrate our attention on the $a \to 0$ limit, and derive the expressions for $a$ and $\sigma$ which replicate best the shape of the $\omega(N,M,S)$ function for each $N$ and $S$, $a(N,S)$ and $\sigma(N,S)$. The mean of the distribution~\eqref{eq:ford} is zero, its variance 
\begin{equation}
\label{eq:ford-var}
v(\sigma,a) \approx \sigma^2 ( 1 - 12 a )
\end{equation}
and the the fourth moment about the mean
\[
\mu_4(\sigma,a) = \int_{-\infty}^\infty \varphi(z) z^4 dz \approx \sigma^4 ( 3 - 96 a ) \,.
\]
Hence, the excess kurtosis is a function of $a$ only
\[
\kappa(a) = \frac{\mu_4(\sigma,a)}{v(\sigma,a)^2} - 3 = - \frac{24 a (1 + 18 a)}{(1-12a)^2} \,.
\]

\begin{figure}[ht!]
\centering\includegraphics[width=\linewidth]{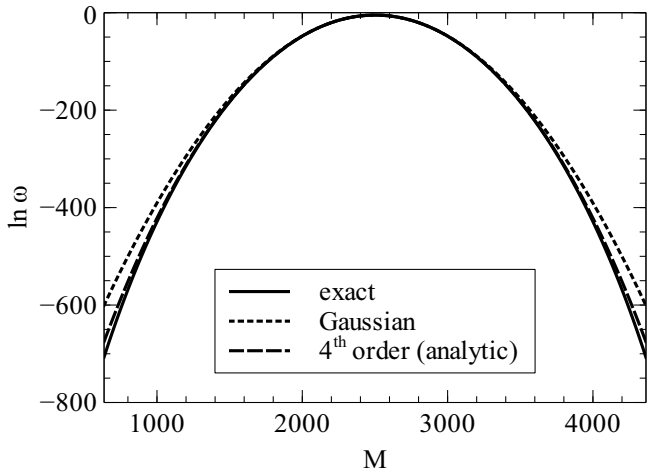}
\caption{Logarithms of exact and approximate values of $\omega(N,M,S)$ for $N = 1000$ and $S=5$ (classical particles, equal mode spacings). The analytic fourth-order formula is much closer to the exact calculations than in the case of $N = 50$ (Fig.~\ref{fig:omg}).}
\label{fig:omg1000}
\end{figure}

For a given value of $\Kurt[\rv M] < 0$, we can solve for $a$ such that~\eqref{eq:ford} has the same excess kurtosis, obtaining
\begin{equation}
\label{eq:aNS}
a(N,S) = \frac{\Kurt[\rv M] + \sqrt{1 - 5 \Kurt[\rv M]} - 1}{12(3 + \Kurt[\rv M])} \,.
\end{equation}
Inserting this into~\eqref{eq:ford-var}, we obtain
\begin{equation}
\label{eq:sigmaNS}
\sigma(N,S) = \sqrt{ \frac{\Var[\rv M] (3 + \Kurt[\rv M]) }{ 4 - \sqrt{1 - 5 \Kurt[\rv M]} } } \,,
\end{equation}
and the approximation reads
\begin{equation}
\label{eq:OmgApprx2}
\begin{split}
&\Omega(M,N,S) \approx \Omega(N,\ev[\rv M],S) \times\\ 
&\,\exp \left( - \frac{(M-\ev[\rv M])^2}{2 \sigma(N,S)^2} - a(N,S) \frac{(M-\ev[\rv M])^4}{\sigma(N,S)^4} \right) .
\end{split}
\end{equation}
The variance and kurtosis of $\rv M$ are given by~\eqref{eq:varM} and~\eqref{eq:kurtM}. As shown by the ``4th order (analytic)'' curve in Fig.~\ref{fig:omg}, for $N=50$ the analytic approximations to $a(N,S)$ and $\sigma(N,S)$ work better than the Gaussian approximation, but are not optimal. However, one can expect that their accuracy will grow with $N$ due to decreasing kurtosis. To test this, we plot the exact values and both (Gaussian and fourth-order) analytic approximations for $N=1000$ and $S=5$ in Fig.~\ref{fig:omg1000}.
Indeed, the fourth-order analytic approximation works much better than the Gaussian and matches quite closely the exact calculations.

\subsection{Quantum particles}

Given the values of the mean, variance and kurtosis of $\rv M$, we can use the Gaussian or fourth-order approximation for $N$ independent fermionic or bosonic particles. However, the computation of these quantities is not as easy as for classical particles, because the particles' excitations cannot be identified with independent random variables, due to their quantum nature. For example, given a system of two bosons which can occupy states 0 and 1, the joint probability distribution of their excitations in the microcanonical ensemble is given by the table
\begin{center}
\begin{tabular}{ccc}
\hline
Configuration & Probability \\
\hline
(0,0) & 1/3\\
(1,0) or (0,1) & 1/3 \\
(1,1) & 1/3 \\
\hline
\end{tabular}	
\end{center}
Such a probability distribution cannot be realized by two independent random variables. On the other hand, the mode occupation numbers $n_{st}$ \emph{are} independent, if we assume that the number of particle in the system is not fixed. Let us assume that we have two modes 0 or 1 and that their occupation numbers $n_1,n_2$ are independent and uniformly distributed between 0 and 2. There are 9 combinations of occupation numbers, each having probability 1/9. If we now impose the constraint $n_1 + n_2 = 2$, we obtain a conditional distribution of occupation numbers given in the table below (where we have used the curly braces to distinguish particle configurations from mode occupations): 
\begin{center}
\begin{tabular}{ccc}
\hline
Occupations & Probability \\
\hline
$\lbrace 2,0 \rbrace$ & 1/3\\
$\lbrace 1,1 \rbrace$ & 1/3 \\
$\lbrace 0,2 \rbrace$ & 1/3 \\
\hline
\end{tabular}	
\end{center}
The distribution in the second table is identical to the one in the first table, with $\lbrace 2,0 \rbrace$ corresponding to configuration (0,0), $\lbrace 1,1 \rbrace$ to configurations (1,0) or (0,1), and $\lbrace 0,2 \rbrace$ to (1,1). It follows that the sum of states $\Omega(M,N,S)$ for quantum particles is proportional to the probability distribution of the variable $\rv M | \rv N=N$ ($\rv M$ conditional on $\rv N$ equal to $N$), where $\rv M$ is the total excitation of the system ($\sum_{st} E_s n_{st}$), and $\rv N$ (also a random variable now) is the total number of particles ($\sum_{st} n_{st}$). The fact that the independent random variables are now associated with mode occupation numbers and not with particle excitations is a reflection of the fact that the particles themselves are excitations of a matter field in Quantum Field Theory, while the constraint on the number of this excitations ($\rv N = N$) is a conservation law characterizing massive particles.

To approximate the distribution of $\rv M | \rv N=N$ analytically, we calculate its mean, variance and kurtosis, and then use the fourth-order approximation~\eqref{eq:OmgApprx2}. The calculation of the above moments is done recursively. Let $\rv M_t$ denote the total excitation of first $t$ quantum modes. (An excitation mode $E_s$ with degeneracy $g_s$ is equivalent to $g_s$ quantum modes, each with the same excitation value.) Let $\mu_k(n,t) = \ev[ \rv M_t^k | \rv N = n]$ and $p(n,t) = P(\rv N = n)$ conditioned on particles occupying first $t$ quantum modes. We have
\begin{equation}
\label{eq:pnt}
p(n,t) \sim \sum_{n'=0}^{\eta(1,n)} p(n-n', t-1) \,,\ \sum_{n=0}^N p(n,t) = 1
\end{equation}
and
\begin{equation}
\begin{split}
\label{eq:muknt}
&\mu_k(n,t) = \sum_{m=0}^n \ev\left[(E_t m + \rv M_{t-1})^k | \rv N = n-m \right]\\
&\qquad \times p(n-m,t-1)\,.
\end{split}
\end{equation}
For example,
\[
\mu_1(n,t) = \sum_{m=0}^n [E_t m + \mu_1(n-m,t-1) ] p(n-m,t-1) \,.
\]
In general, we can calculate $p(n,t)$ from $p(n,t-1)$ and $\mu_k(n,t)$ from $\mu_k(n,t-1)$. The starting conditions are $\mu_k(n,0) = 0$ and $p(n,0) = \delta_{1,0}$. Unlike in the case of classical particles, the above approximation requires a recursive numerical calculation. However, this calculation is less onerous than the one in~\eqref{eq:OmgRec1}, as we do not loop over $M$ values. Additionally, our exact method of calculating the moments $\mu_k(n,t)$ is much simpler than the commonly used non-exact approach, which requires the approximation of the microcanonical density of states using the canonical or grand canonical density of states and saddle-point approximation~\cite{lax:1955,agrawal:1997}.

\begin{figure}
\centering\includegraphics[width=\linewidth]{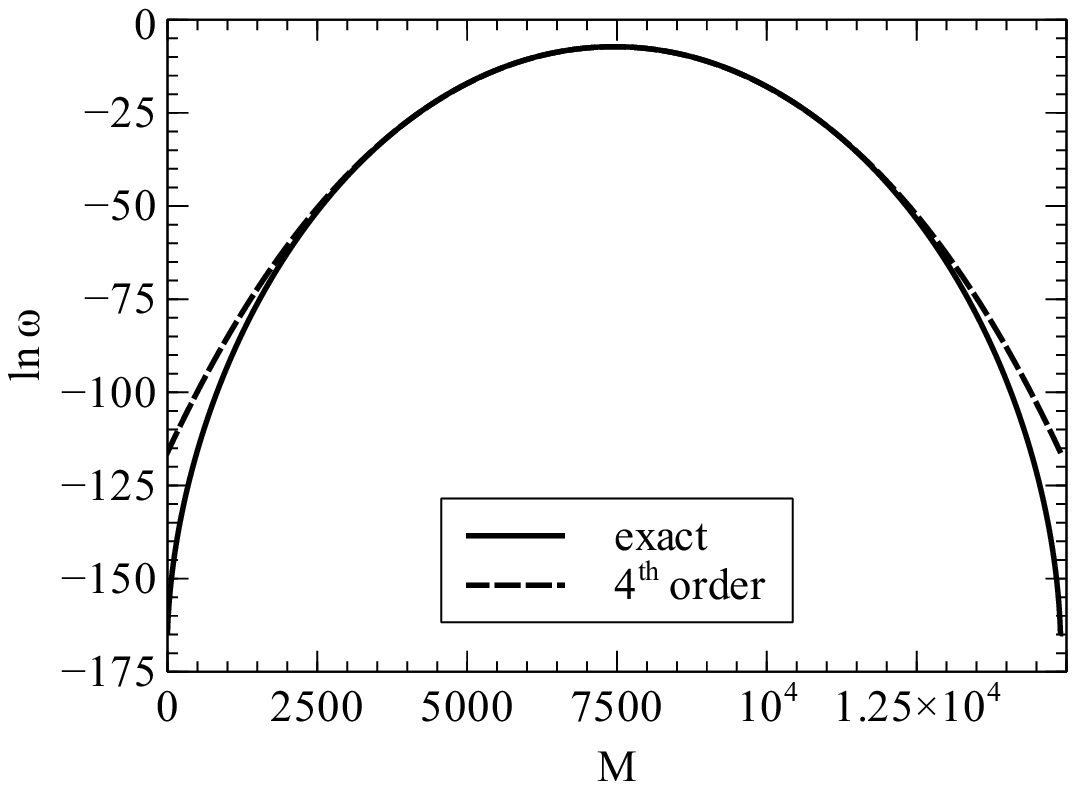}
\caption{Logarithms of exact and approximate sum of states for bosons in a 1D harmonic potential, with $N = 100$ and $150$ modes.}
\label{fig:ht1d-bose}
\end{figure}

\begin{figure}
\centering\includegraphics[width=\linewidth]{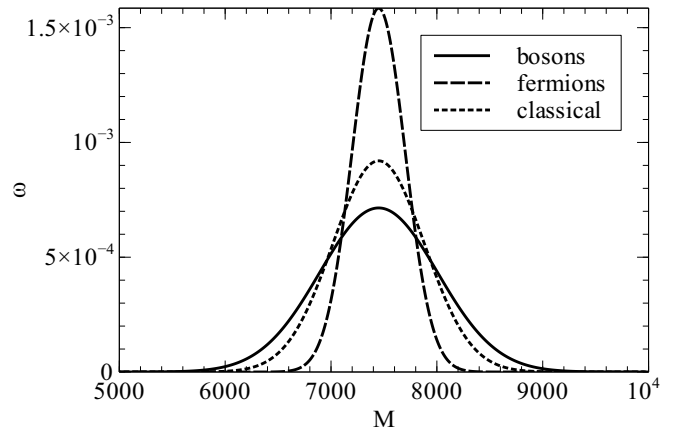}
\caption{Approximate sum of states for bosons, fermions and classical particles in a 1D harmonic potential, with $N = 100$ and $150$ modes.}
\label{fig:ht1d-cmp}
\end{figure}

We test the approximation in the case of a 1D harmonic oscillator potential, for which $g_s \equiv 1$ and $E_s = s$ (equal mode spacings). In the case of bosons, the approximation agrees well with exact results (Fig.~\ref{fig:ht1d-bose}).
Similar results were obtained for fermions. Figure~\ref{fig:ht1d-cmp} compares the results of the approximation for different particle statistics, showing how important it is to include the effect of quantum indistinguishability and especially the Pauli principle (which is omitted by e.g.~the Bethe formula used to describe the sum of states of nuclear spins~\cite{Bethe:1936,agrawal:1997}). In the tests for non-equal mode spacings the proposed approximation is at least as good as the Gaussian one.

\subsection{Non-homogeneous excitation spectra}
\label{sec:non-homogeneous}

The formula~\eqref{eq:OmgApprx2} describes a symmetric sum of states function. Thus, our approximation will not work too well for systems with strongly non-homogeneous excitation spectra, where the dependence of the sum of states on $M$ is strongly skewed, e.g.~with $E_s = s^n$ for $n\ge2$. This can be improved at the cost of making the approximate density of states non-analytic, and using the distribution
\begin{equation}
\label{eq:skew-ford}
\varphi_{\text{skew}}(z) \sim \begin{cases}
\exp \left( - \frac{(z-z_0)^2}{2\sigma_1^2} - \frac{a_1 (z-z_0)^4}{\sigma_1^4} \right) & z < z_0 \\
\exp \left( - \frac{(z-z_0)^2}{2\sigma_2^2} - \frac{a_2 (z-z_0)^4}{\sigma_2^4} \right) & z \ge z_0
\end{cases}
\end{equation}
instead of~\eqref{eq:ford}. Together, the parameters $z_0$, $a_{1,2}$ and $\sigma_{1,2}$ control the mean variance, skewness (third central moment divided by the third power of the standard deviation) and kurtosis of $\varphi_{\text{skew}}$, which can be expressed in a closed form. Linearizing the dependence of these moments on the above parameters leads to a more complex system of equations. One is thus forced to obtain the parameters' values by multidimensional nonlinear fitting, losing the simplicity of the approximation~\eqref{eq:OmgApprx2}. On the other hand, taking into account the skewness of the exact density of states allows to model better the excitation spectra with strongly non-homogeneous densities, such as $E_s = s^2$ (square potential well).
\begin{figure}[ht!]
\centering\includegraphics[width=\linewidth]{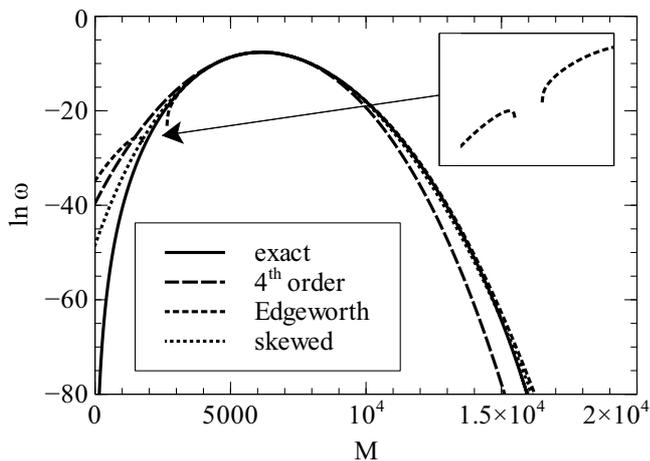}
\caption{Logarithms of exact and approximate values of $\omega(N,M,S)$ for $N = 50$ and 20 modes (classical particles, quadratic mode spacings). The inset shows the range of $M$ values for which the Edgeworth expansion fails to preserve the positivity of the sum of states.}
\label{fig:skewed}
\end{figure}
This example is presented in Fig.~\ref{fig:skewed} and compared with another skewness-sensitive approximation, the Edgeworth expansion (see e.g.~\cite{Blinnikov:1998}). The latter uses Chebyshev-Hermite polynomials to approximate a non-Gaussian probability distribution, and thus does not guarantee the positivity of probability density (Fig.~\ref{fig:skewed}, inset). In our comparison we have used the second-order Edgeworth expansion, which uses the same set of distribution moments as our fourth-order approximation. It is given by the formula
\begin{equation*}
\omega(y) = \frac{e^{-y^2/2}}{\sqrt{2\pi}\sigma} \left( 1 + \frac{H_3(y) \kappa_3}{6 \sigma^3} + \frac{H_4(y) \kappa_4}{24 \sigma^4} + \frac{H_6(y)\kappa_3^2}{72\sigma^6} \right) ,
\label{eq:edgy}
\end{equation*}
where $H_r$ is the $r$-th Chebyshev-Hermite polynomial, $y = (M - \kappa_1) / \sigma$, $\kappa_k$ is the $k$-th cumulant of the approximated distribution and $\sigma = \sqrt{\kappa_2}$. The proposed skewness-sensitive approximation is as close, or closer (for low values of $M$) to the exact results as the Edgeworth expansion, and does not generate negative values.

\section{Physical applications}

In this section we discuss a number of physical problems in which our exact iterative formulas and analytic approximations can be applied. 
 
The iterative expression~\eqref{eq:OmgExpl} can be used with arbitrarily discretized or real-valued mode excitations (measured or obtained numerically), such as isolated quantum dots with known number of electrons, often used to realize qubits (see e.g.~\cite{hitachi}). The ``modes'' of a single dot are individual electronic configurations, and the mode excitations are the total energies of electrons inside the dot. A number of such dots can be thus treated as a system of classical independent particles, and its thermodynamic properties described using the iterative formula~\eqref{eq:OmgExpl}. Since it treats every system configuration separately, this formula can also be used outside thermodynamics to calculate quantum properties which depend on the details of particular configurations, such as Hamiltonian or thermal density matrix elements.

The next problem to which our fourth-order approximation can be applied is the entropy of a black hole and its relation to the area of the event horizon~\cite{blackholes,Bhaduri:2004}. It is well known that the appropriate ensemble in which to calculate the black hole entropy is the microcanonical one. This is true even for evaporating black holes, as they are not in the thermodynamic equilibrium~\cite{casadio:2011}. Within the Loop Quantum Gravity theory, which quantizes geometry, entropy as the function of the event horizon area will be proportional to the logarithm of the number of different microstates of quantized area observables which sum up to a given total area. As these observables are distinguishable, we should use the classical particle statistics. The previous calculations of the microcanonical black hole entropy rely on the assumption that the excitation spectrum of the area operator is evenly spaced~\cite{blackholes,Bhaduri:2004}, which is not true~\cite{ashtekar:1996}. Our results allow to extend the analytic approach to capture the full spectrum of the area operator excitations. Similarly, they can be applied to the statistics of photons in an isolated microwave cavity containing a number of distinguishable atoms~\cite{philippe:2001}.

Another problem for which our approximation is useful is the calculation of the nuclear state density~\cite{Bethe:1936,agrawal:1997}, i.e.~the density $p(M)$ of nuclear microstates which realize a particular value $M$ of the $z$ component of the total angular momentum. Bethe approximated $p(M)$ with a Gaussian distribution (using the Central Limit Theorem) and assuming classical statistics for the particles constituting the nucleus~\cite{Bethe:1936}. This was a major drawback of his approach, as the constituent particles of a nucleus are really fermions, and thus the Bethe approximation can be used only for small values of the total angular momentum of the nucleus~\cite{paar:1993}. An improvement of this approximation uses the Edgeworth expansion (see e.g.~\cite{Blinnikov:1998}) matching the cumulants of $p(M)$ calculated either in the canonical or grand canonical ensemble~\cite{agrawal:1997}. As we have stated in Sec.~\ref{sec:non-homogeneous}, the Edgeworth expansion does not guarantee the positivity of probability density. Our method is free from this problem. 
Although both methods capture higher moments of the density of states, the latter has the advantage that the resulting entropy $k_B \ln \Omega$ is a polynomial function of the total excitation, and thus is easier to treat analytically. Additionally, the calculation of matched moments is not relegated to other statistical ensembles, but is done consistently in the microcanonical one---see~\eqref{eq:pnt} and~\eqref{eq:muknt}.
\begin{figure}%
\centering\includegraphics[width=\linewidth]{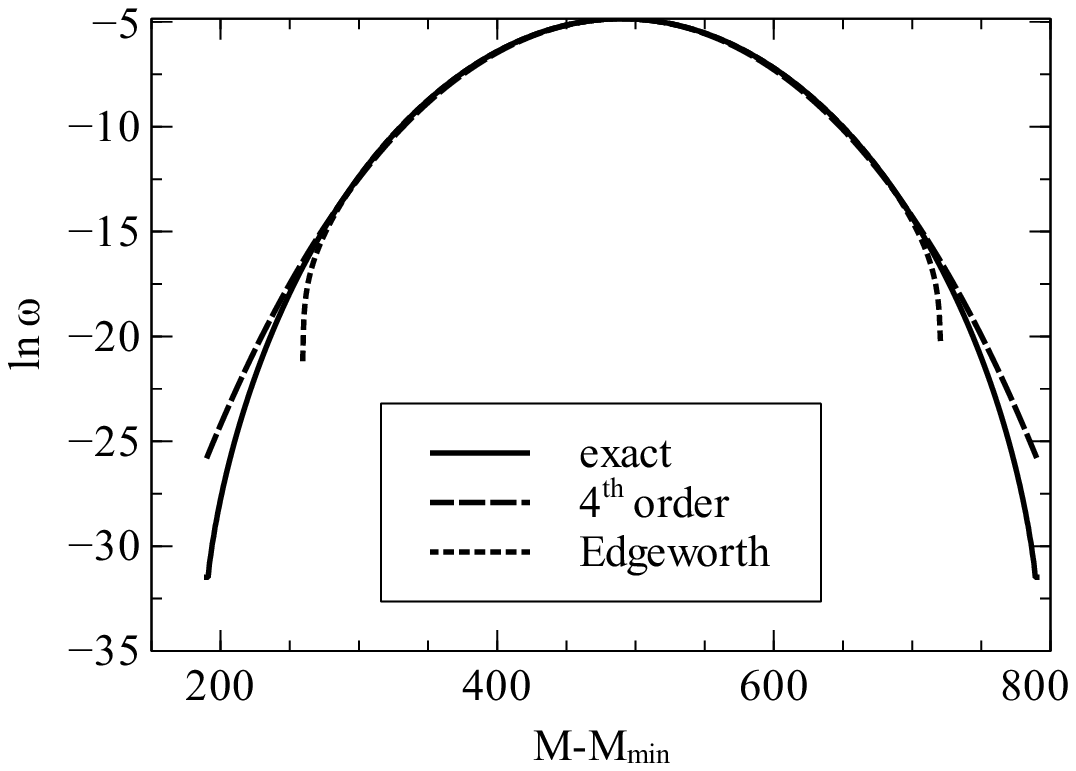}%
\caption{Comparison of the non-Gaussian approximations (our fourth-order and the second-order Edgeworth expansion) for the case of Fermi statistics and equal mode spacings, with $N=20$ and 50 modes. }%
\label{fig:fermcmp}%
\end{figure}
Consequently, as Fig.~\ref{fig:fermcmp} shows, our approximation achieves a better fit to the exact calculations as compared with the second-order Edgeworth expansion (which utilizes the same set of moments as our approximation, but is nevertheless more complicated). The presented results are for Fermi statistics only, but the comparison for the other two gives similar results.

Finally, we note that the free energy $F = E - T\mathcal{S} = E - k_B T \ln \Omega(M)$ becomes within our approximation a polynomial function of the total excitation $M$, which is particularly convenient when using the Ginzburg-Landau formalism (see e.g.~\cite{rilwen}). Because it preserves the higher order $M^4$ term in the exponent, our method can be particularly useful when modeling systems containing small numbers of particles, in which the deviations from the Gaussian limit become stronger. It has this advantage also over the Edgeworth approximation, which (although taking into account higher moments) has only the $M^2$ term in the exponent of the density of states, and therefore does not lead to any free energy terms of higher than quadratic order.

\section{Summary}

We have derived two exact expressions for the marginal or cumulative sum of states of classical, bosonic or fermionic particles, which are useful in practical calculations. Exploiting a probabilistic analogy, we have then obtained an analytic fourth-order approximation to the normalized sum of states (density of states) $\omega$, which captures its variance and kurtosis. The approximation can be extended to an analytic-numerical scheme which also matches the skewness of $\omega$ and is better suited to handle excitation spectra with strongly non-homogeneous densities. As shown numerically, for all three particle statistics our analytic formula is superior to the commonly used Gaussian one, which captures only the first and second moment of the density of states, and also to the Edgeworth expansion. Additionally, our approach employs a simple exact method of calculating the moments of the microcanonical density of states, which requires less computational effort than the commonly used approximations. To prove the above, we have tested our method numerically and discussed its applicability to various physical systems.


\end{document}